# Two-band effects in transport properties of $MgB_2$

M.Putti[1], V.Braccini[1], E.Galleani[1], F.Napoli[1], I.Pallecchi[1], A.S.Siri[1], P.Manfrinetti[2], A.Palenzona[2]

[1]INFM/CNR, Dipartimento di Fisica, Via Dodecaneso 33, 16146 Genova, Italy
[2]INFM, Dipartimento di Chimica e Chimica Industriale, Via Dodecaneso 31, 16146 Genova, Italy

**Abstract**
We present resistivity and thermal conductivity measurements on bulk samples, prepared either by a standard method or by a one-step technique. The latter samples, due to their high density and purity, show residual resistivity values as low as 0.5 µΩ cm and thermal conductivity values as high as 215 W/mK, higher than the single crystal ones. Thermal and electrical data of all the samples are analysed in the framework of the Bloch-Gruneisen equation giving reliable parameter values. In particular the temperature resitivity coefficient, obtained both from resistivity and thermal conductivity, in the dirty sample comes out ten time larger than in the clean ones. This result supports the hypothesis of ref. [1] that π and σ bands conduct in parallel, prevailing π conduction in clean samples and σ conduction in dirty samples .

## 1 Introduction

Since the discovery of superconductivity in Magnesium diboride, this compound appeared to be a "simple" metal where most of the electronic properties follow in a first approximation the basic transport laws of a metal where electron-phonon interaction is dominant. We remind that magnetoresistance follows the Kohler law [2,3], resistivity follows the Bloch-Grüneisen relationship [4,5], Seebeck effect follows the Mott law [4][6]. A further analysis is evidencing the important role in the transport properties of the peculiar band structure of this compound. Two kinds of bands having quite different character contribute to the transport [7,8]: two σ bands, which are hole-type and two-dimensional (2D), and two π bands, which are electron-type and three-dimensional (3D). Calculations showed that σ bands are more strongly coupled with phonon modes and in particular with the $E_{2g}$ mode. A recent paper [1] suggests that just due to the different parity of the two bands, interband impurity scattering turns out to be negligible and the σ and π bands behave as two separate conduction channels in parallel. In this paper we discuss some implications that this fact does have on transport properties, namely electrical and thermal conductivity.

## 2 Sample preparation

Bulk samples were prepared as usual by grinding, pressing into pellets and sintering $MgB_2$ powders previously synthesized from the pure elements. Such samples showed a density of about 1.8 g/cm$^3$. Denser (up to 2.4 g/cm$^3$, 90% of the theoretical density), cleaner and harder cylinder shaped samples have been prepared by a single step method [9], using a similar technique as earlier work [2] [10]. Amorphous or crystalline B and Mg, put in Ta crucibles welded under argon and closed in quartz tubes under vacuum, were heated up to 950°C. Using this process samples were synthesizes with isotopically 99.5% enriched $^{11}$B. In figure 1 SEM (Scanning Electron Microscopy) image of an enriched $^{11}$B sample prepared by this "one-step" method is shown. The image shows a network of well connected grains (1-3 µm large). An X rays spectrum of the same sample is shown in fig. 2 . All the lines associated with the Mg$^{11}$B$_2$ phase are present, no extra peaks due to the presence of free Mg , MgO, are detected.

Three kind of specimens were prepared for physical measurements: one sample obtained by traditional sintering (MGB-TS), one sample prepared by one-step method directly in bulk shape from crystalline boron (MGB-1S) , one sample, prepared by one-step method using enriched $^{11}$B (MGB11-1S). The samples were cut in the shape of parallelepiped bar (1-2×2-3×12 mm$^3$).

The resistivity measurements were performed using a standard four-probe technique. The thermal conductivity was measured using a steady state flux method using a heat flux sinusoidally modulated at low frequency (ν=0.003-0.01 Hz ). Under these conditions the thermal conductivity is extracted as $\kappa = J(\nu)/\nabla T(\nu)$, where $J(\nu)$ is the heat flow provided at the frequency ν and $\nabla T(\nu)$ is the temperature gradient oscillating at the frequency ν. The gradient applied to the sample was varied from 0.1 to 1 K/cm. Seebeck effect was measured simultaneously with the thermal conductivity providing a precise determination of the critical temperature. For the measurements of κ we estimate a sensitivity of 0.5% and accuracy of 2%.

## 3 Resistivity measurements

Resistivity measurements of the three samples from 30 to 300 K are shown in fig. 3. The main differences from sintered to one-step samples are the resistivity values which in MGB-TS are one order of magnitude higher than in MGB-1S and MGB11-1S. This large difference is not reflected by the critical temperature values which are nearly the same in the three samples (see tab. 1) and not even by the residual resistivity ratios (RRR=ρ(300K)/ρ(40K)) which changes only from 3 to 15.

The temperature dependence of the resistivity curves was compared with the function:

$$\rho(T) = \rho_0 + (m-1)\rho'\theta\left(\frac{T}{\theta}\right)^m J_m\left(\frac{\theta}{T}\right) \qquad (1)$$

$$J_m\left(\frac{\theta}{T}\right) = \int_0^{\theta/T} \frac{x^m dx}{(e^x - 1)(1 - e^{-x})} \qquad (2)$$

where $\rho_0$ is the residual resitivity, θ is the Debye temperature, ρ' the temperature coefficient of resistivity for T>θ, and m=3-5. In the range from 40 to 100 K all the curves are well fitted by a power law $\rho(T) = \rho_0 + const \cdot T^m$ with m≅3. This behaviour is well established in $MgB_2$ [3]. Thus the three curves were fitted with eq. (2) taking m=3. The best fitting curves obtained with the parameter values listed in tab. 2 are reported in fig.3 as continuous lines. They reproduce the experimental data with great accuracy. As you can see the θ value does not change too much from sample to sample (1050-1240K), in fair agreement with the Debye temperature obtained from other resistivity measurements [5], even though rather lower values has been obtained from heat capacity measurements [10]. The sintered sample has a $\rho_0$ value as high as 39 μΩcm which can be due partly to grain boundary resistance [11], while the one-step samples show quite low $\rho_0$ values, in particular MGB11-1S has a value as low as 0.55 μΩcm. The excellent purity of the enriched [11]B samples has been emphasised also in ref. [12] and mainly it can be ascribed to the very good quality of the Eagle-Picher enriched [11]B.

Also the ρ' coefficients, vary largely from the sintered to the one-step samples, being of the order of $5\times10^{-1}$ μΩ cm $K^{-1}$ for the first and $5-7.7\times10^{-2}$ μΩ cm $K^{-1}$ for the latters.

## 4 Thermal conductivity measurements

Thermal conductivity measurements of the three samples from 10 K to room temperature are shown in fig. 4. The three samples show quite different curves both in value and in behaviour.

The outstanding quality and high density of the one-step samples is evident from thermal transport properties as well. Indeed, these samples have thermal conductivities more than one order of magnitude higher than those prepared by conventional method (MGB-TS), where grain boundaries give the main contribution to thermal resistance [11]. In particular MGB11-1S exhibits a thermal conductivity as large as 215 W/Km at 65 K which is even larger than that of a single crystal [13], proving the excellent purity and density of this sample.

The thermal conductivity in a metal is given by the sum of the electron thermal conductivity, $\kappa_e$, and the phonon thermal conductivity, $\kappa_p$ and in order to analyse the data of fig. 4 it is necessary to estimate the relative weight of these terms. An estimation can be given by considering the effective Lorenz number $L_{eff}=\kappa T\rho$. In fact, this quantity, that the Wiedemann-Franz law assumes equal to

$L_0 = 2.45 \cdot 10^{-8}$ W$\Omega$K$^{-2}$, in metals where $\kappa_e$ dominates is less than $L_0$ showing a minimum at about $\theta/10$ deeper and deeper with increasing sample purity [14]; in alloys where $\kappa_p$ is not at all negligible $L_{eff}$ becomes larger than $L_0$ and the ratio $L_{eff}/L_0 \approx (1 + \kappa_p/\kappa_e)$ gives information on the relative weight of $\kappa_e$ and $\kappa_p$ [15]. In fig. 5 we plot $L_{eff}$ for the three samples from 40 to 300 K. The curves exhibit the typical behaviour of metals with different levels of purity: at low temperatures all the curves merge and approach $L_0$ from the low; they show a minimum around 130 K (~0.1 $\theta$) which is more and more pronounced from MGB11-1S to MGB-TS; finally, the curves increase in different ways: MGB-TS increases linearly and crosses $L_0$ at 200 K, while MGB-1S and MGB11-1S increase tending to $L_0$ from the low. Thus we can estimate that $\kappa_p$ is 20% of $\kappa$ above 200 K in MGB-TS which means $\kappa_p \sim$ 1-2 W/mK. $\kappa_p$ should be nearly the same for the three samples, depending mainly on the grain dimensions which are not so different in all the samples. Finally, considering that below 200 K $\kappa_p$ has to decrease as $(T/\theta)^3$ we can conclude that the thermal conductivity of the three samples is dominated by electrons.

This result allows us to analyse the thermal conductivity data in term of the electron contribution only. This can be done in the same framework in which electrical conductivity was analysed. We can write:

$$W_e = W_e^p + W_e^i \qquad (3)$$

where, for the Matthiessen's rule, the thermal resistance $W_e = 1/\kappa_e$ is the sum of the thermal resistivity for scattering with phonons, $W_e^p$, and for scattering with impurities, $W_e^i$. Following ref. [14] we can write:

$$W_e^i = \frac{\rho_0}{L_0 T} \qquad (4)$$

$$W_e^p = \frac{4\rho'\theta}{L_0 T}\left(\frac{T}{\theta}\right)^5 J_5\left(\frac{\theta}{T}\right)\left\{1 + \frac{3}{\pi^2} n_a^{2/3}\left(\frac{\theta}{T}\right)^2 - \frac{1}{2\pi^2}\frac{J_7(\theta/T)}{J_5(\theta/T)}\right\} \qquad (5)$$

where $n_a$ is the number of electrons per unit cell and per spin in a given band, and the functions $J_m$ are given by eq. (2); the Debye temperature $\theta$, the residual resistivity $\rho_0$, and the temperature coefficient $\rho'$ are the same as in eq (1). In the temperature range of our interest (T<300K) the more important term in the right side bracket is the one proportional to $T^{-2}$. Thus we can make the following approximation:

$$W_e^p \approx \frac{4\rho'\theta}{L_0 T}\left(\frac{T}{\theta}\right)^3 J_5\left(\frac{\theta}{T}\right)\frac{3}{\pi^2}\left(\frac{n_a}{2}\right)^{2/3} = \frac{C}{L_0 T}\left(\frac{T}{\theta}\right)^3 J_5\left(\frac{\theta}{T}\right)$$

where $C = 4\rho'\theta\frac{3}{\pi^2}n_a^{2/3}$ is a constant.

The best fitting procedure is performed from 40 K to 200 K and the curves obtained with the parameter values listed in tab. 2 are reported in fig.5 as continuous lines. The theoretical curves fit the experimental data in the normal state in excellent way for all the samples and this is astonishing considering the very different shape of the curves. The θ and $\rho_0$ values are in fair agreement with those obtained by resistivity measurements. The C coefficients, which vary from sample to sample, can be related to $\rho'$ once $n_a$ has been estimated.

## 5 Discussion

The result we have achieved up to now is that resistivity and thermal conductivity data can be simultaneously analyzed within the same framework, which represents a single-band electron system interacting with three-dimensional acoustic phonons. This framework seems to be quite unrealistic for MgB$_2$ where the presence of more bands is well established. Actually the $T^3$ behaviour in the low temperature resistivity can be considered a signature of the two-bands nature of this compound being common in transition metals [16] where it is related to inter-band scattering processes.

Another aspect needs to be pointed out: from resistivity measurements we find that the temperature coefficient $\rho'$, which is proportional to the strength of the electron-phonon coupling, changes from one sample to one other, increasing monotonously as the residual resistivity increases. This is well evident considering the values in tab. 2 where we can see that $\rho'$ increases of one order of magnitude from the cleanest (MGB11-1S) to the dirtiest (MGB-TS) sample. This violation of the Matthiessen's rule was emphasised in ref. [1] where it was explained taking into account the band disparity of the electronic structure. In this condition the inter-band scattering is inhibited and the resistivity is given by the parallel of the conduction channels of $\pi$ and $\sigma$ bands and it is dominated by the less resistive one. In ref. [1] it was argued that in the cleanest samples $\pi$ contribution prevails, because $\pi$-bands are more mobile and weakly interact with phonons, but, in dirty samples, mainly if the disorder is localized in Mg layers, $\pi$-bands do not conduct, and electrical current is carried only by $\sigma$-bands. Thus, in clean and dirty limits we observe the $\pi$ and $\sigma$ contribution, respectively.

Within this picture we can try to analyze our data assuming that MGB-TS sample is dominated by $\sigma$-bands, and MGB11-1S and MGB-1S samples are dominated by $\pi$-bands.

With this assumption we can try to estimate $\rho'$ by the $C$ coefficients of thermal conductivity. If we assume $n_{a,\sigma} \sim 0.15$ and $n_{a,\pi} \sim 0.36$ [17], we can now estimate $\rho'$ by the $C$ coefficients and values are reported in table 2. The $\rho'$ values obtained by thermal conductivity differ from those obtained by resistivity, by a factor 3, but we confirm that $\rho'$ in MGB-TS is one order of magnitude larger than in MGB11-1S.

The temperature coefficient $\rho'$, according to ref. [18], can be written as:

$$\rho' = \frac{m_{eff}}{ne^2} \frac{2\pi k_B \lambda tr}{\hbar} = \frac{1}{\omega_p^2 \varepsilon_0} \frac{2\pi k_B \lambda tr}{\hbar} \qquad (7)$$

where $m_{eff}$ is the effective mass, $n$ is the electron density, $\lambda_{tr}$ is the electron-phonon coupling constant and $\varpi_p^2 = \frac{ne^2}{m_{eff}\varepsilon_0}$ is the plasma frequency. $\rho'$ can be calculated for $\pi$ and $\sigma$ bands if the electron-phonon coupling and the plasma frequency of $\pi$ and $\sigma$ bands are considered. Thus introducing in the equation (7) the values of $\lambda_{tr}$ and $\omega_p$ listed in tab. 3 (since our samples are poly-crystals we consider the average of $\omega_p^2$) we obtain $\rho_\pi' = 5.9 \times 10^{-2}$ μΩcm/K and $\rho_\sigma' = 3.9 \times 10^{-1}$ μΩcm/K. These values confirm that carriers in the $\pi$ bands are nearly one order of magnitude more mobile than $\sigma$ bands, due both to the lower coupling with phonons and to the larger plasma frequency. Moreover, the theoretical values have the same order of magnitude of the experimental ones and we find that $\rho_\pi'$ agrees very well with $\rho'$ of MGB11-1S and MGB-1S, the cleanest samples, and $\rho_\sigma'$ with $\rho'$ of MGB-TS, the dirtiest sample.

## 6 Conclusion

Resistivity and thermal conductivity measurements performed in bulk and sintered samples support the theoretical predictions of ref. [1] which assumes that $\pi$ and $\sigma$ bands conduct in parallel, prevailing $\pi$ conduction in clean samples and $\sigma$ conduction in dirty samples. In particular, the temperature coefficients $\rho'$ for the cleanest and the dirtiest samples agree quantitatively with those estimated theoretically for $\pi$ and $\sigma$ bands, respectively.

This result appears very astonishing, in fact eq. (5) describes the electron-phonon coupling in the case of 3D carriers and 3D acoustical phonons. Many of these assumption do not apply to $MgB_2$, being a 2D optical mode the one that is coupled with carriers, and being the carriers both 2D ($\sigma$ bands) and 3D ($\pi$ bands).

Also, the $T^3$ behaviour in the low temperature resistivity appears very peculiar. In the Bloch-Gruneisen framework it is related to the presence of two kind of bands strictly correlated by inter-band scattering processes, which should not be the case of $MgB_2$.

Finally, the close agreement between theoretical and experimental values seems to open more doubts, than to answer questions. These doubts could be clarified by a careful analysis of the electron-phonon interaction. Anyway, transport measurements in the normal state have proved to be a very powerful means to clarify the role of π and σ bands in presence of disorder.

**Table 1**

| sample | $T_C^{onset}$ (K) | $\Delta T_C$ (K) | RRR |
|---|---|---|---|
| MGB11-1S | 38.7 | 0.2 | 15.3 |
| MGB-1S | 38.9 | 0.3 | 7.1 |
| MGB-TS | 38.7 | 1 | 3.3 |

**Table 2**

| | from resistivity | | | from thermal conductivity | | | |
|---|---|---|---|---|---|---|---|
| sample | $\theta$ (K) | $\rho_0$ ($\mu\Omega cm$) | $\rho'$ ($\mu\Omega cm/K$) | $\theta$ (K) | $\rho_0$ ($\mu\Omega cm$) | $C$ ($\mu\Omega cm/K^2$) | $\rho'$ ($\mu\Omega cm/K$) |
| MGB11-1S | 1220 | 0.55 | $4.6\times10^{-2}$ | 1190 | 0.50 | 12 | $2.7\times10^{-2}$ |
| MGB-1S | 1160 | 2.1 | $7.7\times10^{-2}$ | 1130 | 1.9 | 18 | $4.0\times10^{-2}$ |
| MGB-TS | 1050 | 39 | $4.9\times10^{-1}$ | 970 | 34 | 34 | $1.6\times10^{-1}$ |

**Table 3**

| | $\lambda_{tr}^{(a)}$ | $\omega_p^{a-b\ (b)}$ (eV) | $\omega_p^{c\ (b)}$ (eV) |
|---|---|---|---|
| $\pi$ | 0.56 | 5.89 | 6.85 |
| $\sigma$ | 1.1 | 4.14 | 0.68 |

(a) From ref. [1]; (b) from ref. [19]

**Table captions**
Table 1. Critical temperature, amplitude of the transition and residual resistivity ratio defined as RRR=$\rho$(300 K)/ $\rho$(40 K).
Table 2. Values of the parameters $\theta$, $\rho_0$ and $\rho'$ obtained from resistivity best fit and $\theta$, $\rho_0$ and $C$, from thermal conductivity best fit. $\rho'$ values from thermal conductivity have been obtained from the $C$ parameter assuming $n_a$=0.36 for MGB11-1S and MGB-1S and $n_a$=0.15 for MGB-TS.
Table 3. Theoretical values of the electron-phonon coupling constants, $\lambda_{tr}$, the plasma frequancies in the a-b plane, $\omega_p^{a-b}$, and in the c-direction, $\omega_p^c$ for $\pi$ and $\sigma$ bands.

**Figure captions**
Figure 1. SEM images of an enriched [11]B bulk sample prepared by the "one-step" method (sample MGB11-1S). The length scale of the picture is indicated in the bottom.
Figure 2. X-Rays pattern diffraction of an enriched [11]B bulk sample prepared by the "one-step" method (sample MGB11-1S).
Figure 3. Resistivity measurements from 40 to 300 K of MGB11-1S, MGB-1S and MGB-TS: the best fitting curves obtained with the parameter values listed in tab. 2 are reported as continuous lines.
Figure 4. Thermal conductivity measurements of the three samples from 10 K to 280 K: the best fitting curves obtained with the parameter values listed in tab. 2 are reported as continuous lines.
Figure 5. The effective Lorenz number $L_{eff}/L_0 = \kappa T\rho/L_0$ from 40 to 280 K of MgB11-1S, MGB-1S and MGB-TS.

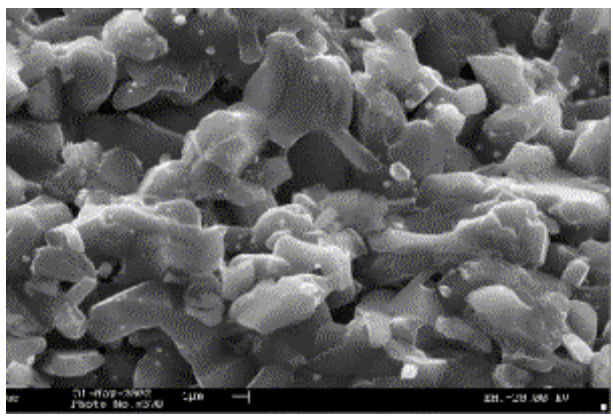

Figure 1

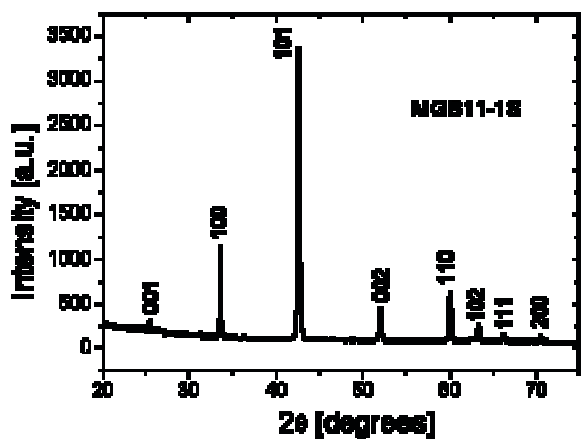

Figure 2

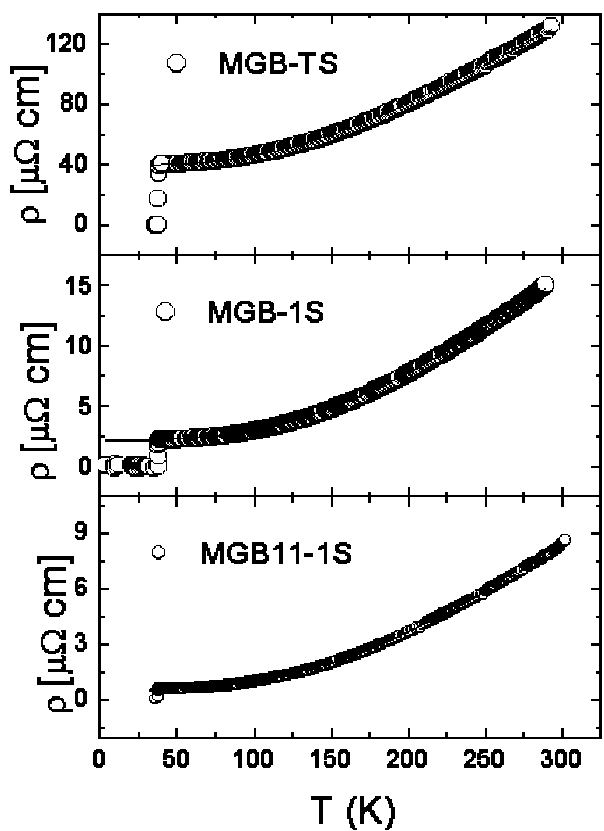

Figure 3

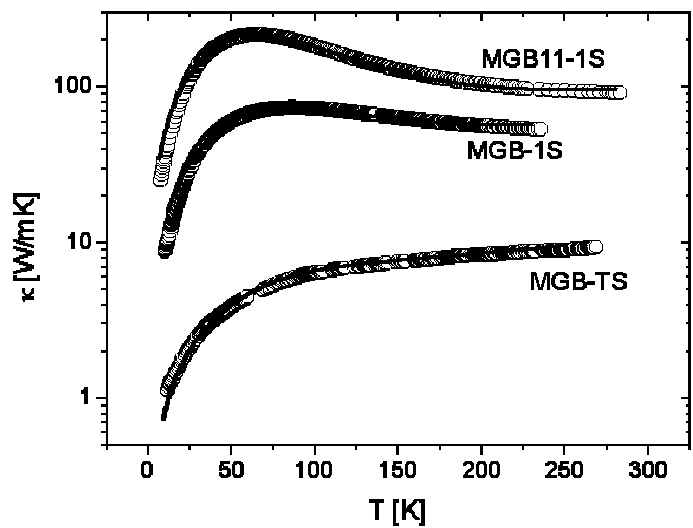

Figure 4

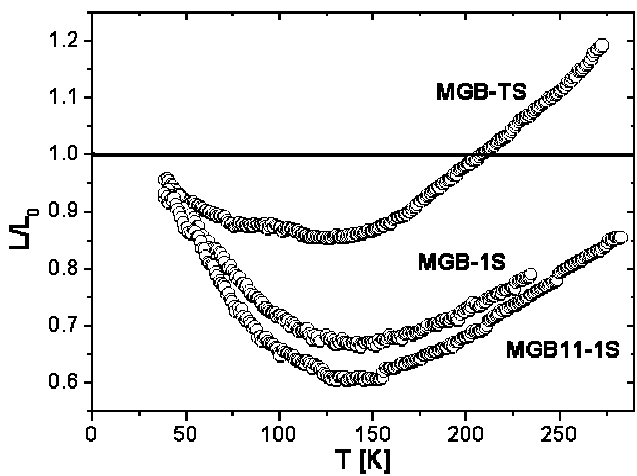

Figure 5